\documentclass[8.5pt,twoside,twocolumn]{article}
\usepackage[super,sort&compress,comma]{natbib} 
\usepackage[version=3]{mhchem}
\usepackage{balance}
\usepackage{times,mathptm}
\usepackage{sectsty}
\usepackage{graphicx} 
\usepackage{lastpage}
\usepackage[format=plain,justification=raggedright,singlelinecheck=false,font=small,labelfont=bf,labelsep=space]{caption} 
\usepackage{fancyhdr}
\usepackage{amsmath}  
\usepackage{bm}     
\newcommand{\nablabf}{\bm{\nabla}}

\begin{document}

\newcommand{\degree}{\ensuremath{^\circ}}

\thispagestyle{plain}
\fancypagestyle{plain}{
\renewcommand{\headrulewidth}{1pt}}
\renewcommand{\thefootnote}{\fnsymbol{footnote}}
\renewcommand\footnoterule{\vspace*{1pt}%
\hrule width 3.4in height 0.4pt \vspace*{5pt}} 
\setcounter{secnumdepth}{5}

\makeatletter 
\renewcommand\@biblabel[1]{#1}            
\renewcommand\@makefntext[1]%
{\noindent\makebox[0pt][r]{\@thefnmark\,}#1}
\makeatother 
\renewcommand{\figurename}{\small{Fig.}~}
\sectionfont{\large}
\subsectionfont{\normalsize} 

\fancyfoot{}
\fancyhead{}
\renewcommand{\headrulewidth}{1pt} 
\renewcommand{\footrulewidth}{1pt}
\setlength{\arrayrulewidth}{1pt}
\setlength{\columnsep}{6.5mm}
\setlength\bibsep{1pt}

\twocolumn[
  \begin{@twocolumnfalse}
\noindent\LARGE{\textbf{Topology optimization of robust superhydrophobic surfaces}}
\vspace{0.6cm}

\noindent\large{\textbf{Andrea Cavalli, Peter B\o ggild, and
Fridolin Okkels\textit{$^{a}$}}}\vspace{0.5cm}

\noindent\textit{\small{\textbf{Received Xth XXXXXXXXXX 20XX, Accepted Xth XXXXXXXXX 20XX\newline
First published on the web Xth XXXXXXXXXX 200X}}}

\noindent \textbf{\small{DOI: 10.1039/b000000x}}
 \end{@twocolumnfalse} \vspace{0.6cm}

  ]

\noindent\textbf{In this paper we apply topology optimization to micro-structured superhydrophobic surfaces for the first time. It has been experimentally observed that a droplet suspended on a brush of micrometric posts shows a high static contact angle and low roll-off angle. To keep the fluid from penetrating the space between the posts, we search for an optimal post cross section, that minimizes the vertical displacement of the liquid-air interface at the base of the drop when a pressure difference is applied.
Topology optimisation proves effective in this framework, showing that posts with a branching cross-section are optimal, which is consistent with several biologic strategies to achieve superhydrophobicity. Through a filtering technique, we can also control the characteristic length scale of the optimal design, thus obtaining feasible geometries}. 
\section*{}


\footnotetext{\textit{$^{a}$Department of Micro- and Nanotechnology, Technical
University of Denmark, DTU Nanotech, Building 345 East, DK-2800
Kongens Lyngby, Denmark}}

\section*{Introduction}

Superhydrophobicity is a remarkable natural phenomenon, recently analysed \cite{marmur2008hygrophilic, deGennes2003capillarity, nosonovsky2007multiscale, mchale2007cassie, patankar2004transition, kusumaatmaja2008collapse} and reproduced artificially \cite{deGennes2003capillarity,tuteja2008omniphobic,bico1999pearl,bico2002wetting,extrand1995inclined} by numerous research groups. Superhydrophobic surfaces show very large static contact angles and small roll-off angles for water, and these properties are usually associated with self-cleaning surfaces.

\begin{figure}[!htb]
\centering
\includegraphics[width=0.45 \textwidth]{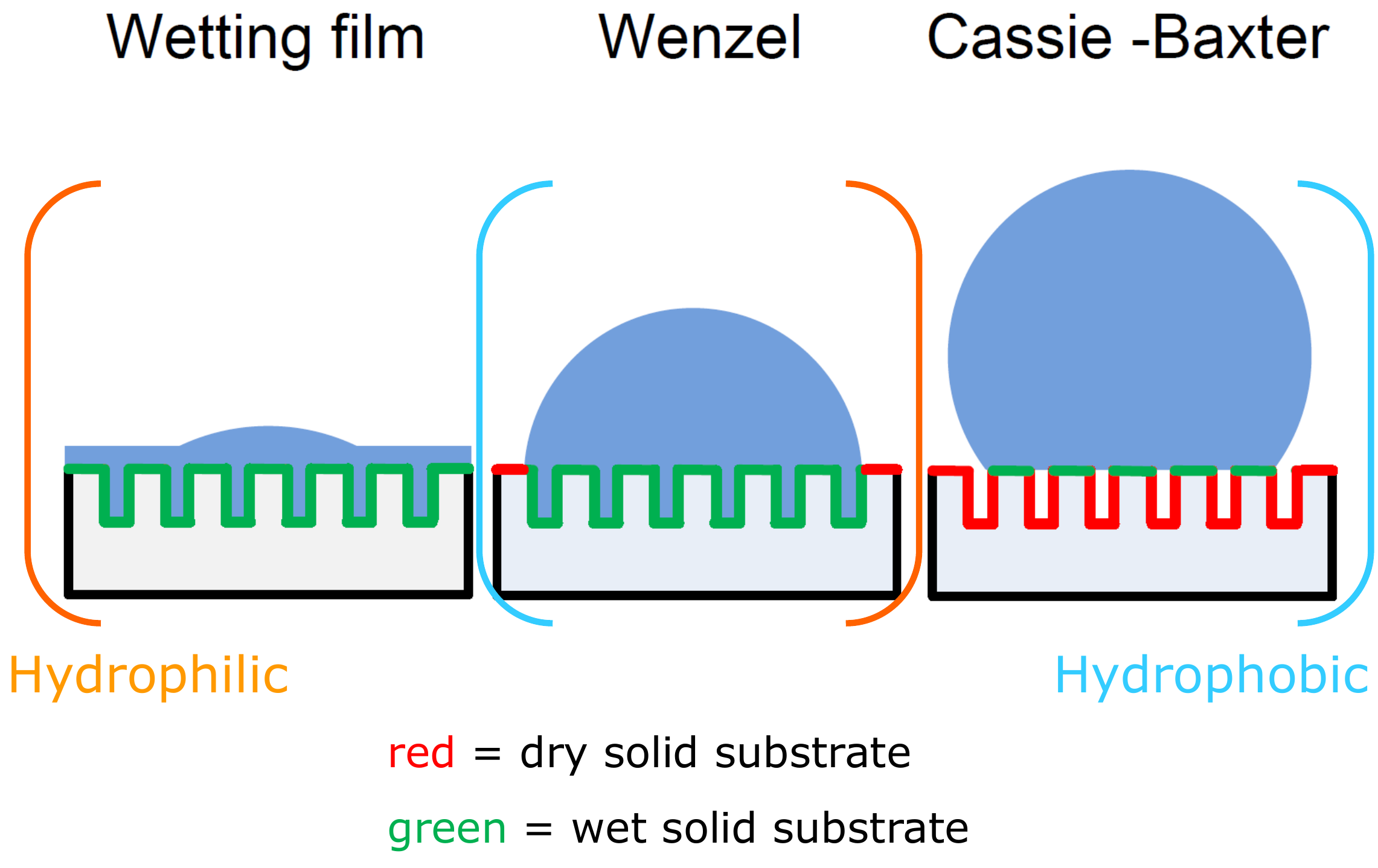}
\caption{Sketch of  the possible equilibrium positions for a drop on a textured surface. The states between orange bracket are accessible for hydrophilic materials, those between blue brackets are accessible for hydrophobic materials.}
 \label{fig:intro}
\end{figure}

A micro- and/or nano-scale texture is usually at the origin of superhydrophobicity \cite{bormashenko2007pigeon, gao2009wetting}.  A drop can reach several different equilibrium states on a textured substrate, as sketched in Fig. \ref{fig:intro}. 
The effective minimum energy configuration depends on the chemical and geometrical properties of the liquid-solid interface. We will now focus on  superhydrophobicity, which is usually associated to the Cassie-Baxter state \cite{cassie1944wettability}. In this configuration, the drop is suspended by the protruding features, so that its base is in contact with a heterogeneous solid-air substrate.
The apparent static contact angle $\theta_{CB}$ , according to Cassie-Baxter theory, is
given by:
\begin{equation}
\label{Cassie}
\cos \theta_{CB} = f_{sl} \cos \theta_Y - (1-f_{sl}),
\end{equation}
 a weighted average between the contact angle for the solid substrate ($\theta_Y$) and for air ($\theta_{air}=180^o$), where $f_{sl}$ represents the wetted solid surface per base area of the drop.

\begin{figure}[!htb]
\centering
\includegraphics[width=0.4 \textwidth]{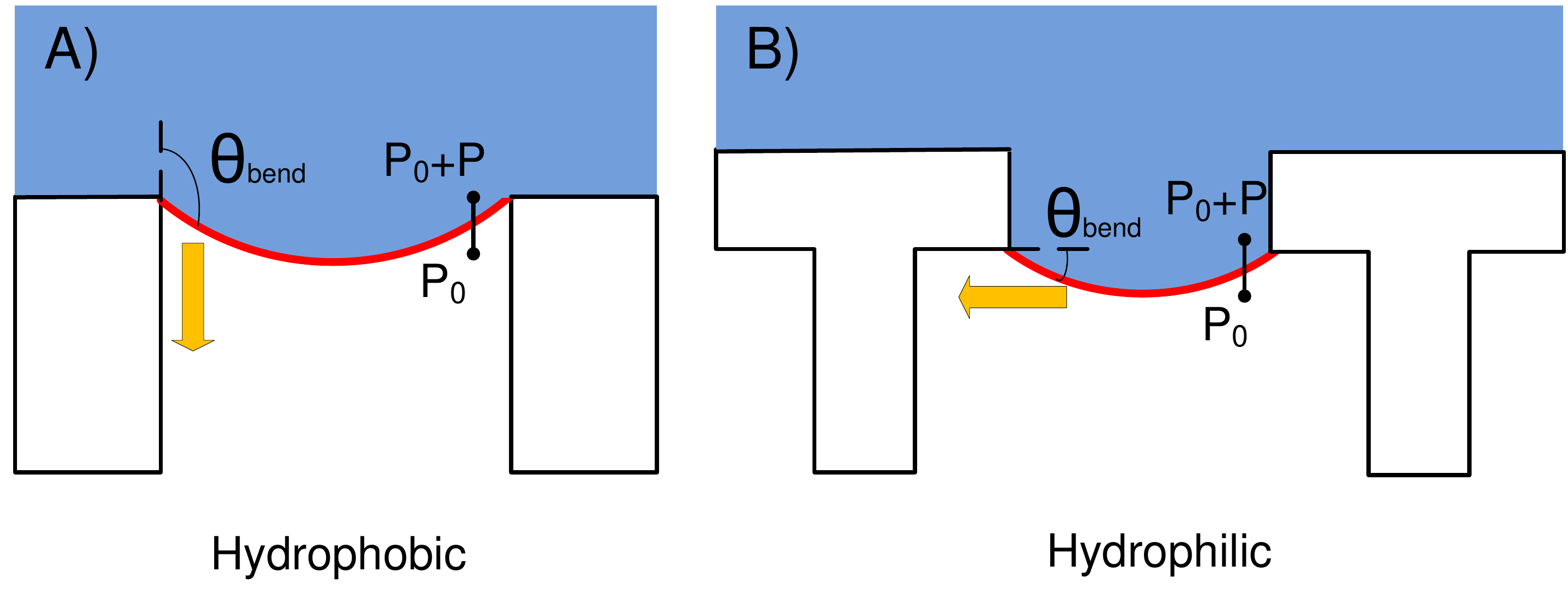}
\caption{A) Interface deformation under applied pressure, for hydrophobic materials. If $\theta_{bend}>\theta_Y$, the contact line slides along the side of the post, in the direction of the yellow arrow B) Analogous setup for hydrophilic materials. If $\theta_{bend}>\theta_Y$, the liquid wets the bottom face of the plate, in the direction of the yellow arrow.}
 \label{fig:bend}
\end{figure}

If a drop in the Cassie-Baxter state is perturbed, for instance if a pressure difference is applied between the drop and the environment, the liquid-air interface will  bulge, and eventually the liquid will begin to flow along the side of the posts when the angle $\theta_{bend}$ (see Fig. \ref{fig:bend}) exceeds the contact angle $\theta_Y$ .

This effect is particularly important for inherently hydrophilic materials, for which a heterogeneous wetting state can be achieved through overhanging structures (see Fig.) \ref{fig:bend}B), even if the global energy minimum will be a Wenzel state \cite{tuteja2008omniphobic} (Fig. \ref{fig:intro}). 
Maximising the robustness of the suspended drop configuration upon applied pressure is therefore fundamental for effective superhydrophobic surfaces.

The research about "Cassie mode" superhydrophobicity has so far been characterized by a strong dichotomy. On one hand, complex hierarchical structures have been fabricated and tested experimentally, but their modelling is hard, since the structures are usually rough and non periodic. \cite{emami2011random} On the other hand, there is an active research for the optimal post shape to achieve a robust Cassie state, which however usually relies on simple shape perturbations to conventional cylindrical or square posts. In this paper, we take a step in bridging this gap, applying the tools of topology optimization.

Topology optimization \cite{bendsoe2003topology} is a structural optimization method with no intrinsic constraint on the topology of the solution, which has been applied in such different fields as structural mechanics \cite{bendsoe2003topology}, photonic crystal design \cite{sigmund2008opttopopt} and microfluidic devices \cite{olesen2006topopt}.
We will here apply it to obtain the texture that minimizes the deformation of the liquid-air interface under applied pressure, thus making the suspended state as robust as possible. 
We will see that this approach generates interesting branching structures, which resemble natural and experimentally tested superhydrophobic structures. However, the symmetry and length scale of the optimal design can be tuned in the numeric optimization procedure, leading to a better understanding and control of such features.

\section*{Modelling and numeric setup}

\begin{figure}[!htb]
\centering
\includegraphics[width=0.38 \textwidth]{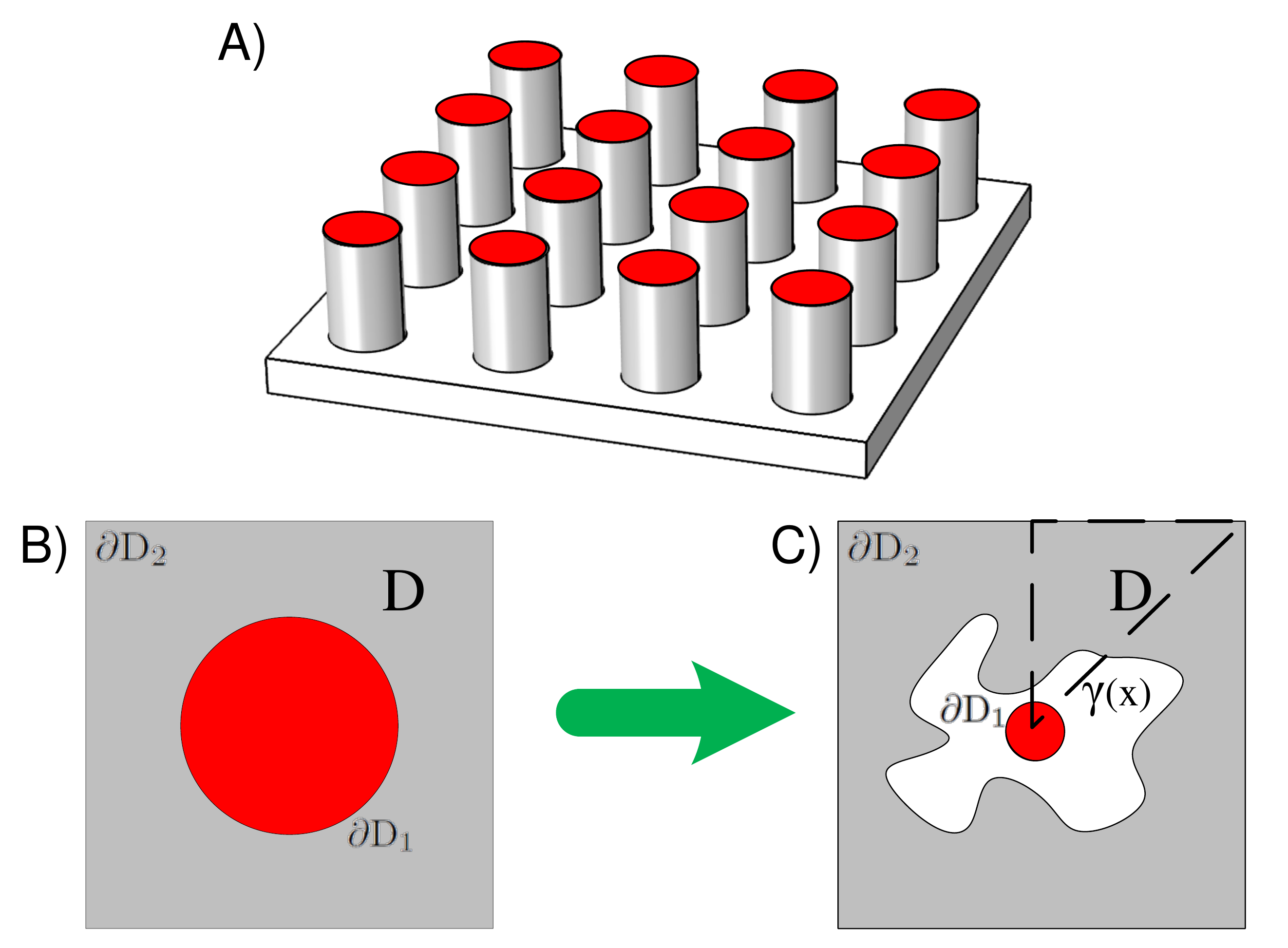}
\caption{A) Sketch of the considered post array. B) Top view of a single post cell, for the basic circular cross section. C) Same view of a single cell, with a variable cross section (white area) around a fixed "nucleus" (red dot) . In the topology optimization procedure, the cross section is not fixed but varies according to the field $\gamma(\vec x)$.  The dashed line in C) shows the reduced computation domain that exploits the symmetry of the cell.}
 \label{fig:sketch}
\end{figure}

In this work we will restrict our analysis to a unit cell for a square array of posts (Fig.\ref{fig:sketch}), neglecting finite size effects at the edge of the drop. We will consider a two dimensional picture, in which the liquid-air interface is flat and suspended on top of the posts (z=0) in the unperturbed configuration, and bulges between the posts to a depth $S(\vec x)$ upon applying a pressure difference $\Delta P$. Such a pressure difference across the liquid-air interface can arise for different reasons, such as the Laplace pressure due to the drop curvature or the pressure upon impact of a drop on the substrate. 
We also introduce non dimensional unit for length $l$, surface tension $\sigma$ and pressure $P$ as follows:

\begin{equation}
\begin{split}
\sigma &= \sigma_{0}\bar \sigma ,\\  
l&=   L_0 \bar l  , \\ \
P&=   \frac{\sigma_{0}}{L_0} \bar P = P_0 \bar P. 
\label{nondimensional}
\end{split}
\end{equation}
Here $L_0$ is the characteristic length of the system, which we will take as the side of the unit cell (typically few $\mu\textrm{m}$), and $\sigma_0$ can be taken as the surface tension of the liquid considered ($\textrm{72.9} \; \textrm{mJ} /\textrm{m}^2$ for water at $\textrm{20}\; \degree\textrm{C}$).
 Moreover, since typically $L_0 << l_c=\sqrt{\frac{\sigma}{\rho g}}$, where $l_c$ is the capillary length for the liquid considered, we can neglect gravity.
 
Let us first consider a simple geometry, such as a cylindrical post ( cross section is shown in Fig.\ref{fig:sketch}B ). The deflection of the liquid-air interface among posts can then be described by the Young-Laplace equation \cite{zheng2005pressure} 
\begin{equation}
\label{bareset}
\begin{cases}
 \nablabf \cdot \left( \frac{\nablabf S(\vec x)}{|\nablabf S(\vec x)|} \right) = \Delta P& \text{on D} \\
S(\vec x)=0& \text{on}  \;\partial \text{D}_1 \\
\nablabf S(\vec x)\cdot \vec n=0& \text{on}  \;\partial \text{D}_2. \\
\end{cases}
\end{equation}

A Dirichlet boundary condition $S(\vec x)=0$ is used at the boundary of the solid structure $\partial \text{D}_1$ to represent that the interface is pinned on the ridge of the post.
 A Von Neumann condition $\nablabf S(\vec x)\cdot \vec n=0$ is applied on the boundary of the unit cell $\partial \text{D}_2$ to account for the symmetry of the post array (in the following, we will also exploit the symmetry of the cell to work only on one eighth of the domain).

For the optimization procedure, we will now slightly modify this setup. We still consider a "solid" support ($S(\vec x)=0$, red dot in Fig.\ref{fig:sketch}C ) in the centre of the domain, but now the post cross section is allowed to change around it, in order to provide an optimal support for the interface.  The distribution of material at point $\vec x$ inside the cell is described by the design variable $\gamma(\vec x)$, a  scalar field which ranges  from 0 (completely solid) to 1 (completely empty) through intermediate values.

 The field $\gamma(\vec x)$ will be coupled to Eqs.\ref{bareset}, leading to the following formulation of the problem:
\begin{equation}
\label{topset}
\begin{cases}
 \nablabf \cdot \left( K(\gamma) \frac{\nablabf S(\vec x)}{|\nablabf S(\vec x)|} \right) = \Delta P& \text{on D} \\
S(\vec x)=0& \text{on}  \;\partial \text{D}_1 \\
\nablabf S(\vec x)\cdot \vec n=0& \text{on}  \;\partial \text{D}_2. \\
\end{cases}
\end{equation}
Where $K(\gamma)$ is defined as:
\begin{equation}
\label{kappa}
K(\gamma)= 1+\frac{(K_{max}-1)\cdot q \cdot(1-\gamma)}{(q+\gamma)}
\end{equation}

Given the form of Eq. \ref{topset} and \ref{kappa}, it is possible to understand the effect of the design variable $\gamma(\vec x)$ on the solution $S(\vec x)$. Where $\gamma(\vec x) = 0$, $K(\gamma) $ is equal to $ K_{max}$, which is fixed to be a large value. The value $\Delta P$ on the right side of Eq.4 becomes then negligible, and the liquid-air interface $S(\vec x)$ does not deform significantly. We therefore recover the "solid" condition $S(\vec x) \simeq 0$.  On the other hand, If $\gamma(\vec x) = 1$ ("empty space"), $K(\gamma) =1$, and we recover the Young-Laplace equation out of the support of the post, as in  Eqs.\ref{bareset}. Intermediate values of $K(\gamma)$ do not have a direct physical interpetation, but are required for a smooth optimization procedure.
The interpolation between these two extreme ranges is controlled by the parameter q in Eq. \ref{kappa}. By choosing a sufficiently small value (here $ 10^{-4}$), it is possible to drive the optimization procedure to give a well defined "solid-empty" binary design. \cite{bendsoe2003topology}.

 This formulation also resembles a 2D optimal heat conduction problem, where $K(\gamma)$ corresponds to the distribution of conducting material \cite{gersborg2006heat}.
 
 We eventually need to define an objective function, i.e. a quantity whose minimization with respect to $\gamma(\vec x)$ will maximise the support to the liquid-air interface.
We choose this quantity, called  $\Phi   \left[ S(\vec x),\gamma(\vec x), \Delta P \right]$, to be the squared integral displacement of the interface (for a given pressure difference $\Delta P$ and material distribution $\gamma(\vec x)$):

\begin{equation}
\label{eq:objective}
\Phi \left[ S(\vec x),\gamma(\vec x), \Delta P \right]   = \int_{D} \, S^2(\vec x) \, dA.
\end{equation}

With this choice, we do not control directly the angle between the interface and the side of the post, which is indeed what would trigger the penetration of the liquid among posts. However, Eq. \ref{eq:objective} is easy to evaluate through the optimization procedure, and its minimization naturally constrains the maximum bending angle of the interface \cite{lobaton2007curvature}, although there might be fluctuations along the post ridge.

At every iteration, the topology optimization code changes the value of $\gamma(\vec x)$ over the domain and evaluates $\Phi   \left[ S(\vec x),\gamma(\vec x), \Delta P \right]$ and the sensitivity $\frac{\delta\Phi}{\delta\gamma(\vec x)}$. We then use this information as input to find the configuration of $\gamma(\vec x)$  that minimizes the objective function $\Phi$, using the method of moving asymptotes (MMA) \cite{svanberg1987mma}. Details on the sensitivity analysis and the implementation of the code can be found in the paper by Olesen et al. \cite{olesen2006topopt}.
We will also introduce a constraint on the maximum solid fraction per unit cell as:
\begin{equation}
\label{constraint}
\int_{D} \, 1 - \gamma(\vec{x}) \, dA \leq f_{sl}.
\end{equation}
Remembering Cassie-Baxter relation $\cos \theta_{CB} = f_{sl} \cos \theta_Y - (1-f_{sl})$ , Eq. \ref{constraint} can conveniently be interpreted as a constraint on the static contact angle shown by a surface patterned in this way.

The specific coupling $K(\gamma)$ we use in Eqs. \ref{topset}  will generate a structure connected to the boundary $\partial \text{D}_1$, which "radiates" the support to the $\gamma  \simeq 0$ regions \cite{gersborg2006heat}. This effectively make our analysis a shape optimization with many degrees of freedom, while the general topology optimization routine we use could as easily generate disconnected topologies. 
 
There are a few reasons for the choice of connected design. First, it is well known that dense and thin posts, ideally down to the nanometer scale, offer increasingly better support to drops in the Cassie-Baxter state \cite{deGennes2003capillarity,zheng2010small}. However, it is perhaps more interesting to optimize the shape of a \textit{single} texture element, which can then be scaled up or down in size according to fabrication and performance constraints. Second, if we are interested in obtaining a hydrophobic behaviour from hydrophilic materials, overhanging structures are required. In this perspective, the central support in our optimisation can be considered as the stem of the post (see Fig. \ref{fig:bend}), while we effectively optimise the cross section of the top plate.
Eventually, we argue that connected structures would show higher mechanical robustness than hair-like features, in particular to buckling and shear loads. This latter property is of great relevance for practical fabrication purposes, since most practical application would include significant stresses for the substrates.\cite{cavalli2012parametric} 

A final remark regards the length scales in the optimal design: at every iteration in the optimisation routine we calculate a smoothed version $\tilde{\gamma}(\vec x)$ of the design variable  $\gamma(\vec x)$, applying a diffusion step \cite{lazarov2011filters}:  

\begin{equation}
\label{diffusion}
L_{diff}^2\nabla ^2 \tilde{\gamma}(\vec x) = \tilde{\gamma}(\vec x)- \gamma(\vec x).
\end{equation}

While calculating the sensitivity, $\tilde{\gamma}(\vec x)$ is then used.
This process allows to control the minimum size of the features appearing in the optimal design. As we will discuss in the next section, without filtering small length scale features would appear in the optimal design, ideally down to the mesh scale. However, these small solid features surrounded by empty space are transformed by the diffusion step in a homogeneous area with intermediate $\gamma(\vec x)$ value, and thus are penalized by the $K(\gamma)$ function, which favours a binary solid-empty solution. The main advantage of this technique is its formulation in terms of a partial differential equation, which relies on the same numeric tools used for  Eqs. \ref{topset}.

The actual implementation of our optimization routine uses a Matlab code, that relies on the commercial software COMSOL  to solve the partial differential equations at every iteration step.

\section*{Discussion of optimized designs}

\begin{figure}[!htb]
\centering
\includegraphics[width=0.35 \textwidth]{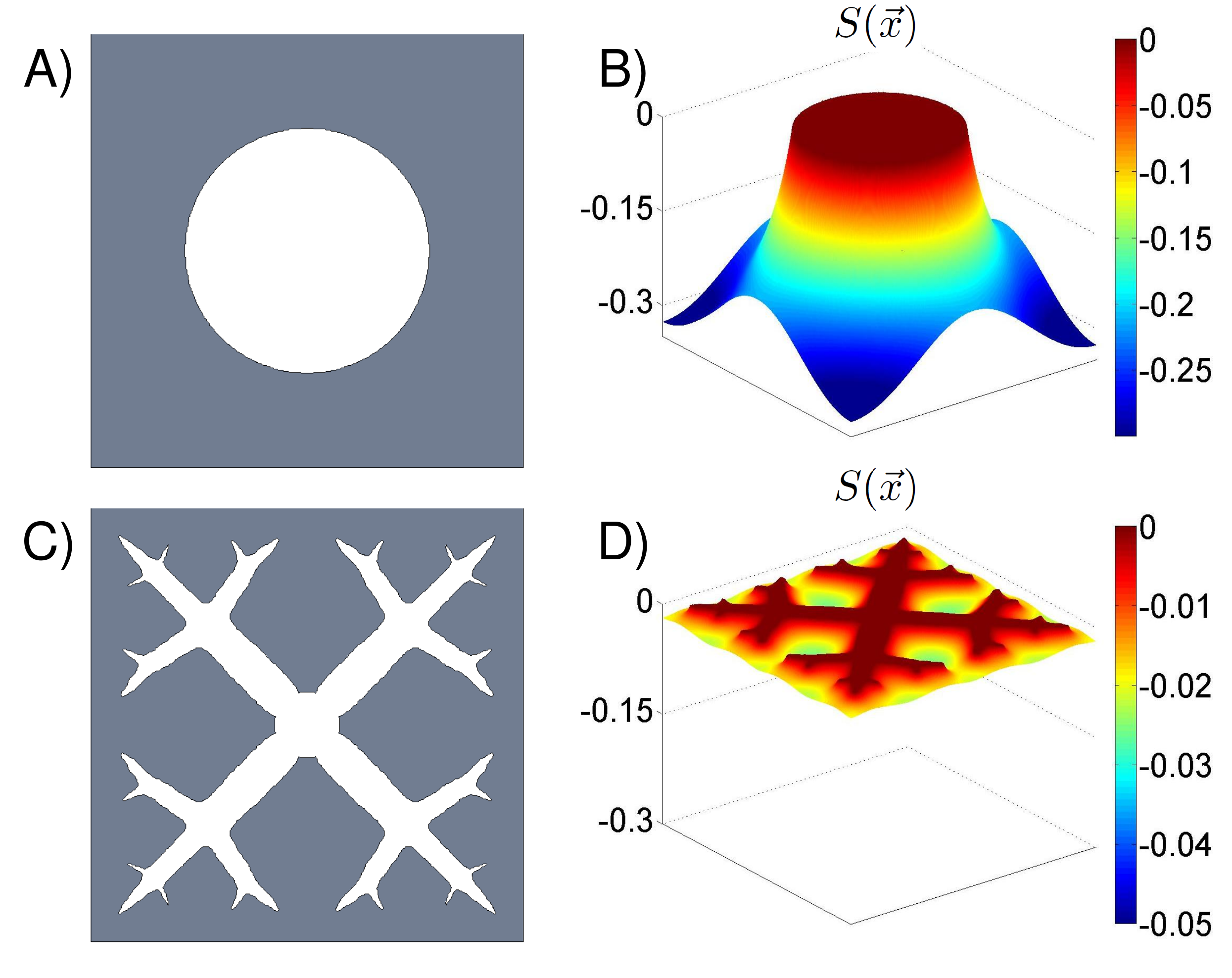}
\caption{A) Top view of a cylindrical post with solid fraction $f_{sl}=0.25$. B) Displacement plot for applied pressure $\Delta P=P_0$. C) Optimized material distribution with solid fraction $f_{sl}=0.25$. D) Displacement plot for the topology optimized design.}
 \label{fig:comparison}
\end{figure}

In the following, the pressure difference acting on the interface has been fixed as $\Delta P = P_0$ .
In Fig. \ref{fig:comparison} We compare the performance of a cylindrical post (A) and an optimized design (C) inside a unit cell. The surface plots displayed on the right (B-D) show the vertical displacement $S(\vec x)$ obtained through Eqs.\ref{bareset}. For both structures, the solid fraction is $f_{sl}=0.25$. It is easy to see the enhanced performance of the topology optimized structure, with the mean displacement reduced by a factor 10.
It is clear that the branching in the optimal structure increases the length of the contact line, where the surface tension acts on the side of the post to balance the effect of the applied pressure difference. This result in a reduction of the interface deformation. However, we think that just choosing a meandering cross section would not improve dramatically the performance. Lobaton and Salamon \cite{lobaton2007curvature}, for instance, considered a simpler sinusoidal perturbation to a circular cross section. While significantly increasing the contact line length, such a shape modification showed modest improvement in the critical pressure value. The added feature of our optimal designs is the convenient placement of the branches, that adjust to the cell shape (here a square unit cell, however analogous solution have been tested for hexagonal lattices) to reduce the size of the gaps between solid features. 
 We therefore argue that the significant reduction in the surface displacement arises from the interplay of optimal location of the main branches and increased  contact line length coming from the secondary branches.

\begin{figure}[!htb]
\centering
\includegraphics[width=0.4 \textwidth]{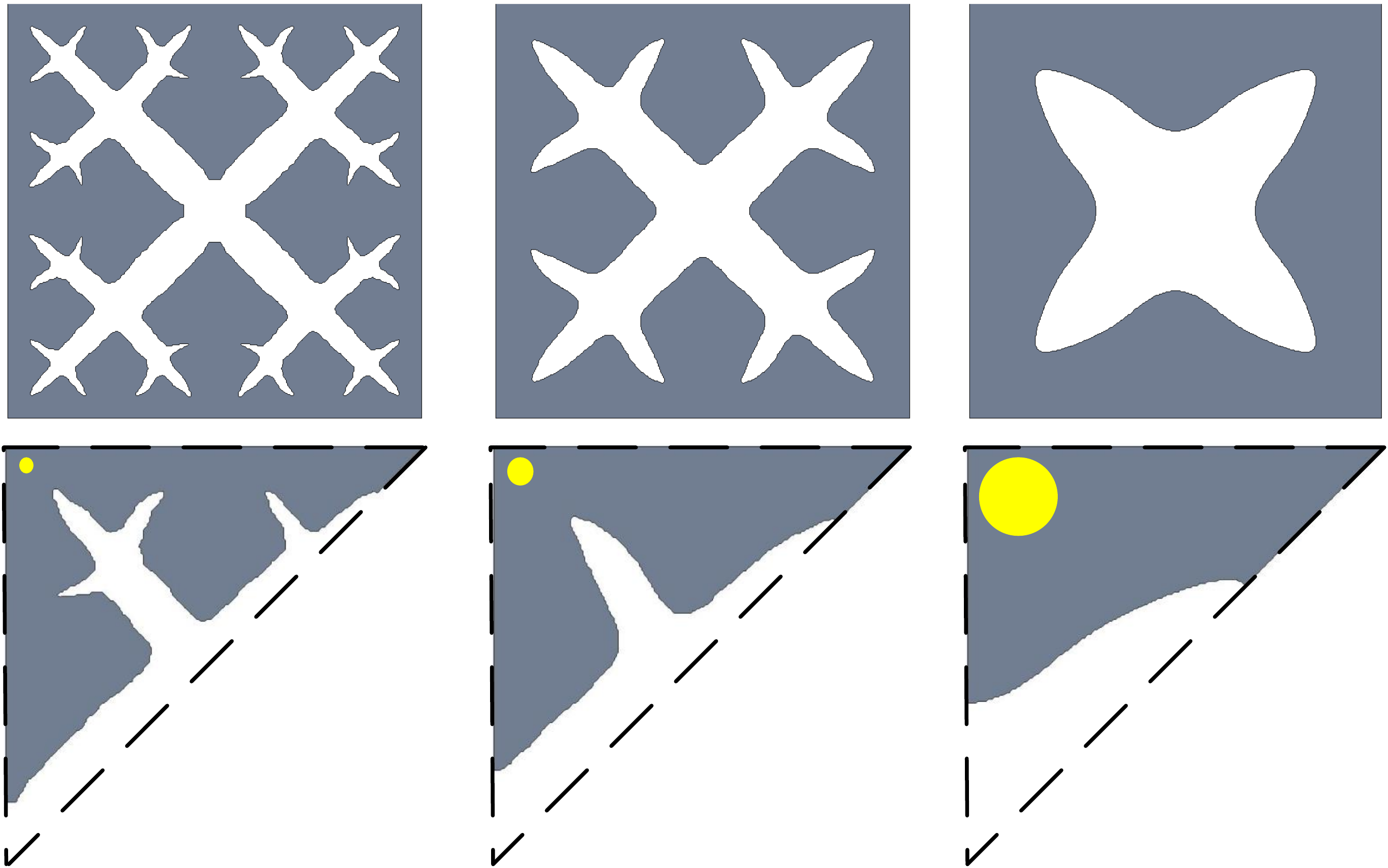}
\caption{Optimal design for $f_{sl}=0.3$ and $L_{diff}$=0.5, 1, 3 times the meshsize $h_{mesh}$. The radius of the yellow dot in each column is equal to $L_{diff}$. }
 \label{fig:filter}
\end{figure}

This physical picture makes it easy to understand the effect of the filtering length  $L_{diff}$ on the optimal design. The designs shown in the upper row of Fig. \ref{fig:filter} were obtained by solving for the domain shown in the bottom row. The yellow dots have a radius equal to $L_{diff}$.

\begin{figure}[!htb]
\centering
\includegraphics[width=0.4 \textwidth]{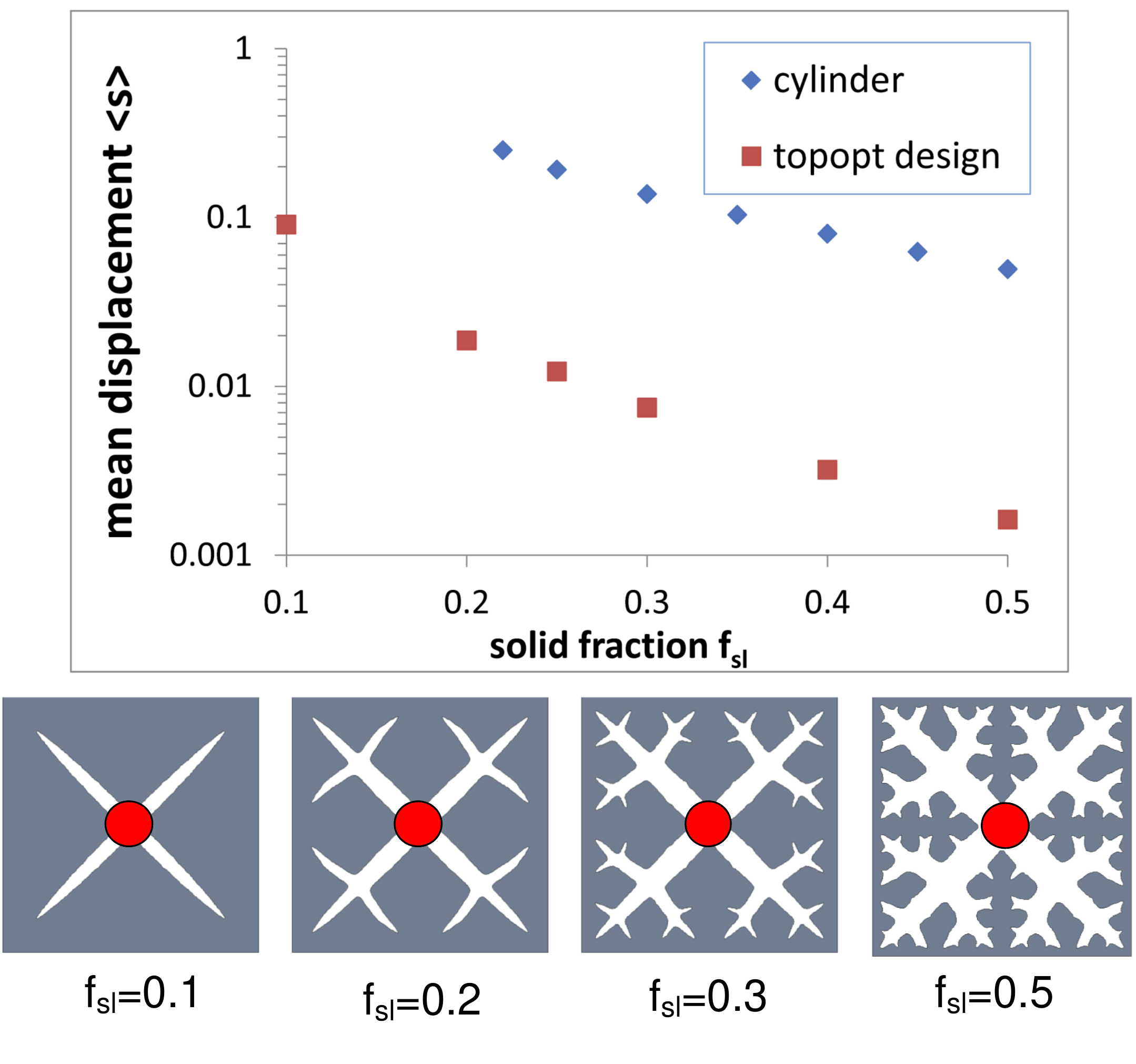}
\caption{Optimal designs as a function of solid fraction $f_{sl}$. The red dot represents the fixed support in the optimisation procedure. In the chart, the mean  vertical displacement of the liquid-air interface is compared for cylindrical posts and topology optimised ones.}
 \label{fig:solidfraction}
\end{figure}

It can be seen that, for any value of $L_{diff}$, the structure branches along the diagonals of the square cell, thus filling the largest gap between two posts. If the resolution is sufficiently fine, further branching appears, with new branches filling the gap among the diagonals. The process continues for even smaller length scales and we get an overall quasi-fractal behaviour.

It is possible to see how the filtering procedure constrains the minimal length scale in the optimal design. This allows to obtain structures suitable for fabrication, i.e. with a feasible amount of branching.

The fractal-like structures resemble several biologic surfaces (such as the lotus leaf), which use analogous (although three dimensional) multi-scale structures to achieve their superhydrophobic properties. 
A three dimensional optimization would be very intensive in terms of computation. It is however possible to complement the suggestions from topology optimization with general knowledge from superhydrophobic surfaces, to get an even more effective texture. Indeed, most artificial and natural superhydrophobic surfaces are characterized by a micron scale texture with superimposed nanometric roughness. The cross sections shown here should be considered an optimal micron scale pattern, over which   nano-grass features can be grown, thus achieving a multi-layer support for the interface (this procedure is currently been considered in collaboration with Nis K. Andersen and Rafael Taboryski, and will be the subject of a future publication).

In Fig. \ref{fig:solidfraction}, we eventually analyse the dependence of the mean interface displacement $\langle s \rangle = \sqrt{\frac{1}{D} \int_{D} \, S^2(\vec x) \, dA}$ on the solid fraction $f_{sl}$ for a fixed filter length $L_{diff}=0.75\; h_{mesh}$, where $h_{mesh}$ is the characteristic mesh size.
An increasing branching for larger solid fraction is clearly seen in the optimal designs, which results in a better support for the interface. In the chart we compare the mean displacement for the optimal design to the displacement for a post of circular cross section and same $f_{sl}$. We can see that the optimised design always performs better than the simple circular cross section, and even more so for large solid fractions, which is again a consequence of the higher degree of branching in the optimised configuration.

\section*{Conclusion and outlook}

In this paper we applied topology optimisation to the stability of superhydrophobic surfaces. We found that this technique is very effective for the task. Branching structures are found to be optimal to support hydrostatic pressure for a Cassie-Baxter state, in a two dimensional analogy to natural structures. We also analysed the effect of a solid fraction constraint on the optimal design, as well as the use of a PDE filter to obtain designs suitable for fabrication.
Further work will include the fabrication and characterization of such optimised microtextured surfaces. Preliminary fabrication results obtained at DTU Nanotech suggest that the optimal shapes can be reproduced with a high degree of precision using common lithographic techniques. A further step will be to use a cost effective procedure, such as injection moulding, to produce the same designs.

This research is funded by the NanoVation consortium. The authors thank Kristian E. Jensen and Rafael J. Taboryski for useful suggestions and discussions.

\bibliography{bib_branches}
\bibliographystyle{rsc}

\end{document}